\documentclass[11pt,twoside]{gsp}
\usepackage{graphicx}
\newcommand{\re}{\mathrm{e}}

\newcommand{\rd}{\mathrm{d}}

\newtheorem{theorem}{Theorem}

\def\aut#1#2{
{\small
\noindent
\parbox[t]{6.5cm}{#1}
 \hfill
\parbox[t]{6.5cm}{#2}
}
}

\begin{document}

\title{Geometry of membranes}
\author{Z. C. Tu}

\date{}

\maketitle

\comm{Communicated by Boris Konopeltchenko}\

\begin{abstract}
This review reports some theoretical results on the Geometry of membranes. The governing equations to describe equilibrium configurations of lipid vesicles, lipid membranes with free edges, and chiral lipid membranes are derived from the variation of free energies of these structures. Some analytic solutions to these equations and their corresponding configurations are also shown.

\end{abstract}

\label{first}

\section{Introduction}
Membranes are very crucial to living organisms. They are the barriers of cells and ensure cells to be relatively isolated individuals but still able to exchange some materials between the inner sides and outer surroundings through specific ways due to the fancy properties of membranes. Membranes usually consist of lipid bilayers mosaicked various kinds of proteins. They are also the key factors to determine shapes of some kinds cells. In particular, the biconcave discoidal shape of red blood cells is regarded as a result of minimizing the free energy of membranes under the area and volume constraints \cite{Canham70,helfrich73} because red blood cells have no complex inner structures. The equilibrium configurations of membranes have attracted much attention of mathematicians and physicists \cite{LipowskyN91,Seifert97ap,oybook,Tu2008jctn,PowersRMP10}. A membrane is thought of as a 2-dimensional (2D) smooth surface in Euclidean space ${\mathbb E}^3$ because its thickness is much smaller than its lateral dimension. The first step to investigate configurations of membranes is constructing a free energy functional by consideration of symmetry. Then the governing equations to describe the equilibrium configurations can be derived by variation of the free energy with some constraints. The next task is seeking for solutions to satisfy the governing equations and comparing the results with typical experiments.

In this review, we will present some purely theoretical results on Geometry of membranes. For simplicity, we merely focus on structures of lipid membranes and only select the theoretical problems that both physicists and mathematicians are interested in. The governing equations to describe equilibrium configurations of lipid structures are derived. Several solutions to these equations and their corresponding geometries are also investigated. The rest of this review is organized as follows. In
Sec.~\ref{sec-prelim}, we give a brief introduction to preliminary in mathematics and physics including surface theory and variational method based on moving frame, and Helfrich's model of lipid bilayer. In Sec.~\ref{sec-vesicles}, we derive a shape equation that describes equilibrium configurations of lipid vesicles --- closed lipid bilayers. Then we discuss some analytic solutions and their corresponding configurations which include surfaces of constant mean curvature, torus, biconcave discoid, cylinder-like vesicles, and so on. In Sec.~\ref{sec-openmem}, we investigate a lipid membrane with free edge(s). The shape equation and boundary conditions describing equilibrium configurations of the membrane are derived. Then we discuss the compatibility between the shape equation and boundary conditions, and verify five theorems of non-existence. In Sec.~\ref{sec-chiralm}, we construct the free energy functional of chiral lipid membranes in terms of symmetric argument and then derive the governing equations to describe their equilibrium configurations by variational method. Some analytic solutions and their corresponding configurations are also shown. In the last section, we give a brief summary and a list of related open questions.

\section{Preliminary in Mathematics and Physics\label{sec-prelim}}

In order to continue our discussion, we first introduce several key mathematical and physical concepts and tools that will be used in the following sections.

\subsection{Surface Theory Based on Moving Frame}

\subsubsection{Moving Frame Method}

\begin{figure}[pth!]
\centerline{\includegraphics[width=7cm]{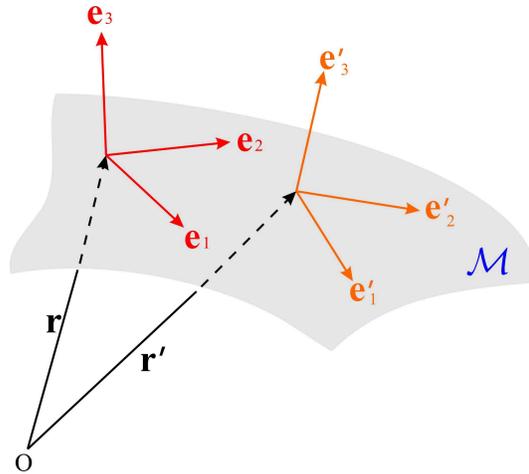}}
\caption{The moving frame on a surface.\label{sframefig}}
\end{figure}

A membrane can be regarded as a smooth orientable surface embedded in ${\mathbb E}^3$. The properties of surface such as mean curvature and Gaussian curvature determine the shape of membrane. As shown in Fig.~\ref{sframefig}, each point on surface $\mathcal{M}$ can be represented a position vector $\mathbf{r}$. At that point we construct three unit orthonormal vectors
$\mathbf{e}_1$, $\mathbf{e}_2$, and $\mathbf{e}_3$ with $\mathbf{e}_3$ being the normal vector of surface $\mathcal{M}$ at point $\mathbf{r}$. The set of right-handed orthonormal triple-vectors $\{\mathbf{e}_1,\mathbf{e}_2,\mathbf{e}_3\}$ is called a frame at point $\mathbf{r}$. Different points on the surface have different vectors $\mathbf{r}$, $\mathbf{e}_1$, $\mathbf{e}_2$, and $\mathbf{e}_3$, thus the set $\{\mathbf{r};\mathbf{e}_1,\mathbf{e}_2,\mathbf{e}_3\}$ is called a moving frame.

Let us imagine a mass point that moves from position $\mathbf{r}$ to its neighbor position $\mathbf{r}'$ on the surface. The length of the path is denoted by $\Delta s$. Then we can define the differentiation of the frame as
\begin{equation}
\rd\mathbf{r}=\lim_{\Delta s\rightarrow
0}(\mathbf{r}'-\mathbf{r})=\omega_1\mathbf{e}_1+\omega_2\mathbf{e}_2,\label{sframer}
\end{equation}
and
\begin{equation}\rd\mathbf{e}_i=\lim_{\Delta s\rightarrow
0}(\mathbf{e}_i'-\mathbf{e}_i)=\omega_{ij}\mathbf{e}_j,\quad
(i=1,2,3)\label{sframee}\end{equation} where $\omega_1$, $\omega_2$,
and $\omega_{ij}~(i,j=1,2,3)$ are 1-forms, and `$\rd$' is the exterior
differential operator\cite{Chernbook}. The 1-form $\omega_{ij}$ is ant-symmetric with respect to $i$ and $j$, that is $\omega_{ij}=-\omega_{ji}$. Here and in the following contents without special statements, the repeated subscripts represent summation from 1 to 3. Additionally, the
structure equations of the surface can be expressed as\cite{Chernbook}:
\begin{align}
\rd\omega_1&=\omega_{12}\wedge\omega_2\notag\\
\rd\omega_2&=\omega_{21}\wedge\omega_1\label{eq-stru}\\
\rd\omega_{ij}&=\omega_{ik}\wedge\omega_{kj}\quad
(i,j=1,2,3),\notag\end{align} and
\begin{equation}
\left(\begin{array}{l}\omega_{13}\\
\omega_{23}\end{array}\right)=\left(\begin{array}{cc}a&b\\
b&c\end{array}\right)\left(\begin{array}{l}\omega_{1}\\
\omega_{2}\end{array}\right),\label{omega13}\end{equation} where
`$\wedge$' represents the wedge production between two differential
forms. The matrix $\left(\begin{array}{cc}a&b\\
b&c\end{array}\right)$ is the representation matrix of the curvature
tensor. Its trace and determinant are two invariants
under the coordinate rotation around $\mathbf{e}_3$ which are
denoted by
\begin{equation}2H=a+c\quad \mathrm{and}\quad K=ac-b^2.\label{relat2HK}\end{equation}
$H$ and $K$ are called the mean curvature and Gaussian curvature, respectively. They determine the shape of the surface
and can be expressed as $2H=-(1/R_1+1/R_2)$ and $K=1/R_1R_2$ with
two principal curvature radii $R_1$ and $R_2$ at each point on the surface.

Now consider a curve on surface $\mathcal{M}$. Its tangent vector is denoted by $\mathbf{t}$.
Let $\phi$ be the angle between $\mathbf{t}$ and $\mathbf{e}_1$ at the same point.
Then the geodesic curvature $k_{g}$, the geodesic torsion $\tau _{g}$, and the normal
curvature $k_{n}$ along the direction of $\mathbf{t}$ can be expressed as\cite{TuJPA04}:
\begin{align}
k_{g}&=(\rd\phi +\omega _{12})/\rd s\notag\\
\tau _{g} &=b\cos2\phi+(c-a)\cos\phi\sin\phi\label{geodisicc}\\
k_{n}&=a\cos^2\phi+2b\cos\phi\sin\phi+c\sin^2\phi,\notag
\end{align}
where $\rd s$ is the arc length element along $\mathbf{t}$. If
$\mathbf{t}$ aligns with $\mathbf{e}_1$, then $\phi=0$,
$k_{g}=\omega _{12}/\rd s$, $\tau _{g} =b$, and $k_{n}=a$. In the principal frame, the geodesic torsion and normal
curvature can be expressed as
\begin{equation}k_{n}=-\frac{\cos^2\phi}{R_1}-\frac{\sin^2\phi}{R_2},~~\tau _{g}=(1/R_1 -1/R_2)\cos\phi\sin\phi.\label{geodisiccprfm} \end{equation}

\subsubsection{Stokes' theorem and related identities}
Stokes' theorem is a crucial theorem in differential geometry, which reads
\begin{equation}\oint_{\partial\mathfrak{D}}\omega=\int_\mathfrak{D} d\omega,\end{equation}
where $\mathfrak{D}$ is a domain with boundary $\partial\mathfrak{D}$. $\omega$
is a differential form on $\partial\mathfrak{D}$. In particular, $\int_\mathfrak{D} d\omega=0$ for a closed domain
$\mathfrak{D}$.

From Stokes' theorem, we can derive several identities listed as follows \cite{TuJPA04,Tu2008jctn}.

(i) For smooth functions $f$ and $h$ on 2D subdomain
$\mathfrak{D}\subseteq \mathcal{M}$,
\begin{align}\int_\mathfrak{D}(f\rd{\ast}\rd h-h\rd{\ast}\rd f)=\oint_{\partial
\mathfrak{D}}(f{\ast} \rd h-h{\ast}
\rd f)\notag\\
\int_\mathfrak{D}(f\rd{\ast}\tilde{\rd}h-h\rd{\ast}\tilde{\rd}f)=\oint_{\partial
\mathfrak{D}}(f{\ast} \tilde{\rd}h-h{\ast}
\tilde{\rd}f)\label{greenident}\\
\int_\mathfrak{D}(f\rd\tilde{\ast}\tilde{\rd}h-h\rd\tilde{\ast}\tilde{\rd}f)=\oint_{\partial
\mathfrak{D}}(f\tilde{\ast} \tilde{\rd}h-h\tilde{\ast}
\tilde{\rd}f),\notag\end{align} where $\ast$ is Hodge star operator satisfying $\ast \omega_1 =\omega_2$ and $\ast \omega_2 =-\omega_1$. $\tilde{\rd}$ and $\tilde{\ast}$ are
generalized differential operator and generalized Hodge star which satisfy
$\tilde{\rd}f= f_1\omega_{13}+f_2\omega_{23}$ and
$\tilde{\ast}\tilde{\rd}f= f_1\omega_{23}-f_2\omega_{13}$ if
$\rd f=f_1\omega_{1}+f_2\omega_{2}$\cite{TuJPA04,Tuvarna}.

(ii) If $\mathbf{u}$ and $\mathbf{v}$ are two vector fields defined on 2D subdomain $\mathfrak{D}\subseteq \mathcal{M}$, then \cite{Tunpublished}
\begin{align}\int_\mathfrak{D}(\mathbf{u}\cdot\rd{\ast}\rd \mathbf{v}-\mathbf{v}\cdot\rd{\ast}\rd \mathbf{u})=\oint_{\partial
\mathfrak{D}}(\mathbf{u}\cdot{\ast} \rd \mathbf{v}-\mathbf{v}\cdot{\ast}
\rd \mathbf{u})\notag\\
\int_\mathfrak{D}(\mathbf{u}\cdot\rd{\ast}\tilde{\rd} \mathbf{v}-\mathbf{v}\cdot\rd{\ast}\tilde{\rd} \mathbf{u})=\oint_{\partial
\mathfrak{D}}(\mathbf{u}\cdot{\ast} \tilde{\rd} \mathbf{v}-\mathbf{v}\cdot{\ast}
\tilde{\rd} \mathbf{u})\label{vgreenident}\\
\int_\mathfrak{D}(\mathbf{u}\cdot\rd\tilde{\ast}\tilde{\rd} \mathbf{v}-\mathbf{v}\cdot\rd\tilde{\ast}\tilde{\rd} \mathbf{u})=\oint_{\partial
\mathfrak{D}}(\mathbf{u}\cdot\tilde{\ast} \tilde{\rd} \mathbf{v}-\mathbf{v}\cdot\tilde{\ast}
\tilde{\rd} \mathbf{u}),\notag\end{align}where the `dot'
represents the inner product of vectors. For simplicity, Eqs.(\ref{greenident}) and (\ref{vgreenident}) are still called Stokes' theorem in this review. They are widely used in the variational process.

Additionally, we can also define the gradient, curl, divergent, Laplace operators, etc. on the surface in terms of the differential operators and Hodge stars. They are summarized as follows \cite{Tu2008jctn,Tupre2007}:
\begin{align}
&(\nabla\times\mathbf{u})\,\rd A=\rd(\mathbf{u}\cdot \rd\mathbf{r}),~(\nabla\cdot \mathbf{u})\,\rd A=\rd(\ast\mathbf{u}\cdot \rd\mathbf{r})\notag\\
&(\tilde{\nabla} \cdot \mathbf{u})\,\rd A=\rd(\tilde{\ast}\mathbf{u}\cdot \tilde{\rd}\mathbf{r}),~(\bar{\nabla} \cdot \mathbf{u})\,\rd A=\rd(\ast\mathbf{u}\cdot \tilde{\rd}\mathbf{r})\notag\\
&\nabla f \cdot \rd\mathbf{r}=\rd f,~\tilde{\nabla} f \cdot \rd\mathbf{r}=\tilde{\rd}f\notag\\
&(\nabla^2f)\,\rd A=\rd\ast \rd f,~(\nabla\cdot\bar{\nabla} f)\,\rd A=\rd\ast\tilde{\rd}f\notag\\
&(\nabla\cdot\tilde{\nabla} f)\,\rd A=\rd\tilde{\ast} \tilde{\rd}f\notag\\
&(\mathbf{u} \cdot\nabla f)\,\rd A=\mathbf{u}\cdot
\rd\mathbf{r} \wedge\ast\rd f\label{basicidenty}\\
&(\mathbf{u} \cdot\bar{\nabla} f)\,\rd A=\mathbf{u}\cdot
\rd\mathbf{r} \wedge\ast\tilde{\rd} f\notag\\
&(\mathbf{u} \cdot\tilde{\nabla} f)\,\rd A=\mathbf{u}\cdot
\rd\mathbf{r} \wedge\tilde{\ast}\tilde{\rd} f\notag\\
&(\nabla^2\mathbf{u})\,\rd A=\rd\ast \rd\mathbf{u},~(\nabla\mathbf{u})\cdot\,\rd\mathbf{r}=\rd\mathbf{u}\notag\\
&(\nabla \mathbf{u}\colon \nabla \mathbf{v})\rd A= \rd \mathbf{u}\dot{\wedge} \rd \mathbf{v}\notag\\
&\mathbf{S}\cdot \rd \mathbf{r}=-\omega_{12}, ~ (\nabla\cdot\mathbf{S})\rd A = -\rd \ast \omega_{12},\notag
\end{align}
where $\rd A=\omega_1\wedge\omega_2$ and $\mathbf{S}$ are the area element and spin connection of the surface, respectively. $\dot{\wedge}$ represents calculating dot production and wedge production simultaneously.

\subsection{Helfrich's Model \label{subs-helfmd}}

In 1973, Helfrich proposed a spontaneous curvature model to describe the free energy of lipid membranes by analogy with a bent box of liquid crystal in Smectic A phase \cite{helfrich73}. The free energy is in fact a functional defined in the space of shapes of membranes, which reads
\begin{equation}F_H = \int_\mathcal{M} \left[ \frac{k_c}{2} (2H +c_0)^2 + \bar{k} K \right] \rd A,\label{helfricheng}\end{equation}
where $\mathcal{M}$, $H$, and $K$ represent the membrane surface, mean curvature, and Gaussian curvature, respectively. $k_c$ and $\bar{k}$ are two bending moduli. The former should be positive, while the latter can be negative or positive for lipid membranes. $c_0$ is called spontaneous curvature, which reflects the asymmetrically chemical or physical factors between two leaves of lipid bilayers.

On the other hand, the spontaneous curvature model can also be obtained from symmetric argument. A lipid membrane can be locally regarded as 2D isotropic elastic entity. Thus the local free energy density $f$ should be invariant under rotational transformation around the normal direction of the membrane surface. In other words, it should be a function of $H$ and $K$ because $H$ and $K$ are the fundamental invariants of the surface under rotational transformation. Up to the second order terms of curvatures, it can be expanded as
\begin{equation}f_c=A_0+ A_1 H + A_2 H^2 +A_3 K,\end{equation}
which can be rewritten as
\begin{equation}f_c=\frac{k_c}{2} (2H +c_0)^2 + \bar{k} K\label{helfrichedens}\end{equation}
with omitting an unimportant constant. This is no other than the integrand in Eq.~(\ref{helfricheng}). Therefore, the spontaneous curvature model is of general significance not only for lipid membranes, but also for other membranes consisting of isotropic materials. The following discussions are mainly based on Helfrich's spontaneous curvature model.

\subsection{Variational Method Based on Moving Frame}

To obtain governing equations that describe equilibrium configurations of lipid membranes, we need to minimize the free energy. That is, we should calculate the variation of functionals defined in the space of shapes of membranes. In this space, each shape might be expressed as a function, but the definitional domain is not a fixed planar domain. It is difficult to deal with this case by using the traditional calculus of variation. Now we will introduce the variational method based on the moving frame developed by the present author and Ou-Yang \cite{Tu2008jctn,TuJPA04,TuPRE03}, which is available to deal with variational problems on surfaces.

Any infinitesimal
deformation of a surface can be achieved by a displacement vector\begin{equation}\delta \mathbf{r}\equiv \mathbf{v}=\Omega_{i}\mathbf{e}_{i}\end{equation} at
each point on the surface, where
$\delta$ can be understood as a variational operator. The frame is also
changed because of the deformation of the surface, which is
denoted as
\begin{equation}\delta\mathbf{e}_i=\Omega_{ij}\mathbf{e}_j\quad (i=1,2,3),\label{varyei}\end{equation}
where $\Omega_{ij}=-\Omega_{ji}~(i,j=1,2,3)$ corresponds to the
rotation of the frame due to the deformation of the surface. From
$\delta \rd\mathbf{r}=\rd\delta\mathbf{r}$, $\delta
\rd\mathbf{e}_j=\rd\delta\mathbf{e}_j$, and
Eqs.~(\ref{sframer})--(\ref{omega13}), we can derive
\begin{align}\delta \omega _{1}&=\rd\mathbf{v}\cdot \mathbf{e}_{1}-\omega _{2}\Omega _{21}=\rd\Omega_{1} +\Omega_2\omega_{21}+\Omega_3\omega_{31}-\omega _{2}\Omega _{21}\notag\\
\delta \omega _{2}&=\rd\mathbf{v}\cdot \mathbf{e}_{2}-\omega
_{1}\Omega _{12}=\rd\Omega_{2} +\Omega_1\omega_{12}+\Omega_3\omega_{32}-\omega
_{1}\Omega _{12}\label{variframe}\\
\delta \omega _{ij}&=\rd\Omega _{ij}+\Omega _{il}\omega _{lj}-\omega
_{il}\Omega _{lj}\notag\\
\rd\mathbf{v}\cdot \mathbf{e}_{3}&=\rd\Omega_3 +\Omega_1\omega_{13}+\Omega_2\omega_{23}=\Omega_{13}\omega_1+\Omega_{23}\omega_2.\notag\end{align} These equations
are the essential equations of the variational method based
on the moving frame. With them as well as
Eqs.~(\ref{omega13}) and (\ref{relat2HK}), we can easily derive
\begin{align}\delta \rd A &=( \mathrm{div\,}\mathbf{v}-2H\Omega _{3}) \rd A \notag\\
\delta (2H) &=[\nabla ^{2}+(4H^{2}-2K)] \Omega_{3}+\nabla (2H)
\cdot \mathbf{v} \label{variAHK}\\
\delta K &=\nabla \cdot \tilde{\nabla}\Omega _{3}+2KH\Omega
_{3}+\nabla K\cdot \mathbf{v}.\notag
\end{align}
Using the above equations (\ref{eq-stru})-(\ref{geodisicc}), (\ref{greenident})-(\ref{basicidenty}), (\ref{variframe}) and (\ref{variAHK}), we can deal with almost all variational problems on surfaces.

\section{Lipid Vesicles\label{sec-vesicles}}

\begin{figure}[pth!]
\centerline{\includegraphics[width=8cm]{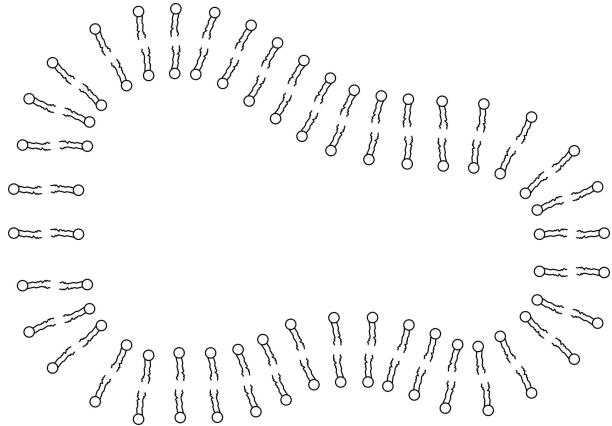}}
\caption{A cartoon of lipid vesicle.\label{figvesicle}}
\end{figure}

Most of lipid molecules are amphiphiles with a hydrophilic head group
and two hydrophobic hydrocarbon tails. When a quantity
of lipid molecules disperse in water, they will assemble
themselves into a bilayer vesicle as depicted in Fig.~\ref{figvesicle} due to hydrophobic forces. In this section, we will theoretically understand various configurations of lipid vesicles.

\subsection{Shape Equation to Describe Equilibrium Configurations}

A lipid vesicle can be represented as a closed surface $\mathcal{M}$. Its equilibrium shape is determined by minimizing Helfrich's free energy (\ref{helfricheng}) under the constraints of constant area and volume because experiments reveal that the area of lipid membranes are almost incompressible and the membranes are impermeable for the solutions in both sides of the membranes. Thus we can introduce two Lagrange multipliers $\lambda$ and $p$ to replace these two constraints, and then minimize the following functional
\begin{equation}F = \int_\mathcal{M} \left[ \frac{k_c}{2} (2H +c_0)^2 + \bar{k} K \right] \rd A +\lambda A +pV ,\label{fegvesicle}\end{equation}
where $\mathcal{M}$, $A$ and $V$ represent the membrane surface, total area of the vesicle and volume enclosed by the vesicle. $\lambda$ and $p$ can also be understood as the apparent surface tension and osmotic pressure of the lipid vesicle.

The Euler-Lagrange equation of functional (\ref{fegvesicle}) can be derived by using the variational method presented in Sec.~\ref{sec-prelim}, which reads \cite{OYPRL87,OYPRA87,TuJPA04}
\begin{equation}\tilde{p}-2\tilde{\lambda}
H+(2H+c_0)(2H^2-c_0H-2K)+2\nabla^2H=0\label{shape-closed}\end{equation}
with reduced parameters $\tilde{p}=p/k_c$ and $\tilde{\lambda}=\lambda/k_c$. This formula is called the shape equation because it describes the equilibrium shapes of lipid vesicles and represents the force balance along the normal direction of membrane surfaces.

\begin{figure}[pth!]
\centerline{\includegraphics[width=8cm]{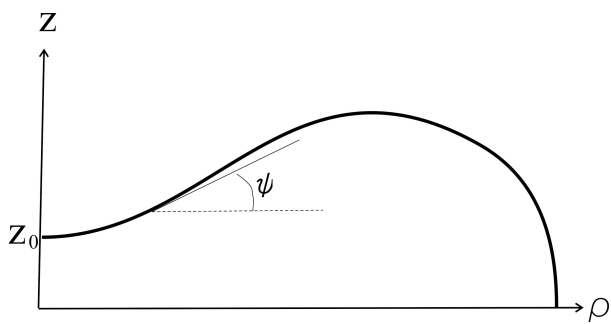}}
\caption{The generation curve for an axisymmetric vesicle.\label{figgencuve}}
\end{figure}

Now let us consider an axisymmetric vesicle generated by a planar curve shown in Fig.~\ref{figgencuve}. Rotate the curve around $z-$ axis and then mirror with respect to the horizontal plane. $\rho$ is the revolving radius, and $\psi$ is the angle between the tangent of the curve and the horizontal plane. That is, $dz/d\rho =\tan \psi$. Through simple calculations, we can obtain
$-2H=h\equiv {\sin \psi }/{\rho}+(\sin\psi)'$, $K={\sin \psi
}(\sin\psi)'/{\rho}$, and $\nabla^2(2H)=-({\cos \psi }/{\rho})(\rho\cos \psi h')'$,where the `prime' represents the derivative with respect to $\rho$. Substituting these relations into Eq.~(\ref{shape-closed}), we derive
\begin{equation}
(h-c_{0})\left(\frac{h^{2}}{2}+\frac{c_{0}h}{2}-2K\right)-\tilde{p}-\tilde{\lambda}
h+\frac{\cos \psi }{\rho}(\rho\cos \psi h')'=0,\label{shape-symmetr}\end{equation}
The equivalent form of this equation is first derived by Hu and Ou-Yang \cite{HuOY93PRE}.
The shape equation (\ref{shape-symmetr}) of axisymmetric vesicles is a third-order differential
equation. Following Zheng and Liu's work \cite{zhengliu93}, we can
transform it into a second order differential equation
\begin{equation}\cos\psi h'
+(h-c_{0}) \sin\psi\psi^{\prime}-\tilde{\lambda} \tan\psi+\frac{2\eta_{0}-\tilde{p}\rho^2}{2\rho\cos\psi}-\frac{\tan\psi}
{2}(h-c_{0})^{2} =0\label{firstintg}\end{equation} with an integral
constant $\eta_{0}$. It is found that the shape equation of axisymmetric vesicles obtained by Hu and Ou-Yang degenerates into that derived by Seifert \emph{et al.} \cite{SBLPRA91} when $\eta_{0}=0$ in Eq.~(\ref{firstintg}) or for vesicles with spherical topology free of singular points \cite{zhengliu93,Podgornikpre95}.

\subsection{Analytic Solutions and Corresponding Configurations}

Now we will show several analytic solutions to shape equations (\ref{shape-closed}) or (\ref{firstintg}) of lipid vesicles that we have known till now. They correspond to surfaces of constant mean curvature, torus, biconcave discoid, and so on. Most of these solutions are found by Ou-Yang and his coworkers. We will see that only sphere, torus, and biconcave discoid can correspond to lipid vesicles. We also recommend the reader to note the classic paper by Konopelchenko \cite{Konop97} who expressed Eq.~\ref{shape-closed} in terms of the inverse mean curvature and density of squared mean curvature, and then found a broad variety of its solutions and corresponding possible shapes.

\subsubsection{Surfaces of Constant Mean Curvature}
\begin{figure}[pth!]
\centerline{\includegraphics[width=11cm]{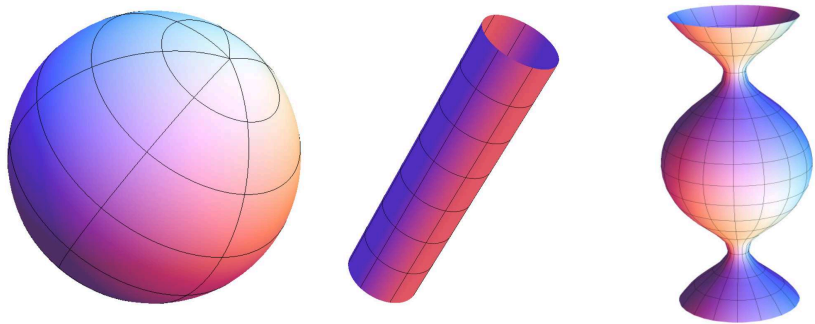}}
\caption{Surfaces of constant mean curvature: sphere (left), cylinder (middle), and unduloid (right).\label{figconstH}}
\end{figure}

Obviously, surfaces of constant mean curvature (including sphere, cylinder, and unduloid shown in Fig.~\ref{figconstH}) can satisfy the shape equation. First, let us consider a spherical surface with radius $R$. Then $H=-1/R$ and $K=1/R^2$. Substituting them into Eq.~(\ref{shape-closed}), we derive
\begin{equation}\tilde{p}R^2+2\tilde{\lambda}
R-c_0(2-c_0R)=0.\label{sphericalbilayer}\end{equation} This
equation gives the relation between sphere radius $R$, spontaneous curvature $c_0$, reduced osmotic pressure $\tilde{p}$, and reduced surface tension $\tilde{\lambda}$.

Next, $H=-1/2R$ and $K=0$ for a cylindrical surface with radius $R$. Then shape equation (\ref{shape-closed}) requires
\begin{equation}2\tilde{p}R^3+2\tilde{\lambda}
R^2-1+c_0^2R^2=0.\label{cylinderbilayer}\end{equation}

For other surfaces of constant mean curvature such as unduloid, $H$ is a constant but $K$ is not a constant.
In terms of shape equation (\ref{shape-closed}), we obtain
\begin{equation}H=-c_0/2\label{constHbilayer}\end{equation}
and $\tilde{p}=-\tilde{\lambda}c_0$.

Note that cylindrical surface and unduloid are in fact not closed surfaces. Alexandrov proved that ``An
embedded surface (no self-intersection) with constant mean curvature in $\mathbb{E}^3$ must be a spherical
surface'' \cite{Alexandrov62}. Thus we can only observe one kind of vesicles of constant mean curvature --- sphere.

\subsubsection{Torus}

\begin{figure}[pth!]
\centerline{\includegraphics[width=9cm]{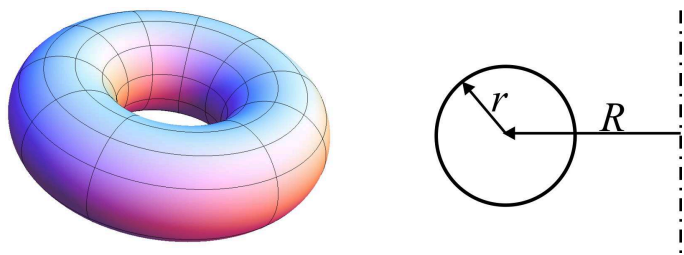}}
\caption{Torus (left) and its generation curve (right).\label{figtorus}}
\end{figure}

As shown in Fig.~\ref{figtorus}, A torus is a revolution surface generated by a circle with radius
$r$ rotating around an axis in the same plane of the circle. The
revolving radius $R$ should be larger than $r$. The
torus can be expressed as vector form
$\{(R+r\cos\varphi)\cos\theta,(R+r\cos\varphi)\sin\theta,r\sin\varphi\}$.
Through simple calculations, we have $2H=-(R+2r\cos\varphi)/r(
R+r\cos\varphi)$ and $K=\cos\varphi/r(R+r\cos\varphi)$.
Substituting them into Eq.~(\ref{shape-closed}), we derive
\begin{eqnarray}
&&\hspace{0.26cm} [(2 c_0^2r^2-4 c_0r+4\tilde{\lambda}
r^2+2\tilde{p}r^3)/{\nu^3}]\cos^3\varphi\nonumber\\
&&+[(5 c_0^2r^2-8 c_0r+10\tilde{\lambda} r^2+6\tilde{p}r^3)/{\nu^2}]\cos^2\varphi\nonumber\\
&&+[{(4 c_0^2r^2-4 c_0r+8\tilde{\lambda} r^2+6\tilde{p}r^3)}/{\nu}]\cos\varphi\nonumber\\
&&+2/{\nu^2}+(c_0^2r^2-1)+2(\tilde{p}r+\tilde{\lambda})r^2=0\label{toruseq}
\end{eqnarray}
with $\nu=R/r$. If $\nu$ is finite, then
Eq.~(\ref{toruseq}) holds if and only if the coefficients of
$\{1,\cos\varphi,\cos^2\varphi,\cos^3\varphi\}$ vanish. It follows
$2\tilde{\lambda} r=c_{0}( 4-c_{0}r)$,
$\tilde{p}r^{2}=-2 c_{0}$ and
\begin{equation}\nu=R/r=\sqrt{2}.\end{equation}
That is, there exists a lipid torus with the ratio of its two
generation radii being $\sqrt{2}$ (called $\sqrt{2}$ torus by Ou-Yang \cite{oypra90}), which was confirmed in the
experiment \cite{MutzPRA91}. It is also found that nonaxisymmetric tori \cite{Seiferttorus} constructed from conformal transformations of $\sqrt{2}$ torus also satisfy the shape equation.

To check the consistency, we need also verify that the above toroidal solution can indeed satisfy shape equation (\ref{firstintg}) of axisymmetric vesicles. It is not hard to see $\sqrt{2}$ torus can be generated from a curve expressed as
\begin{equation}
\sin\psi = (\rho/r) \pm \sqrt{2}.\label{toruseqsym}
\end{equation}
Substituting it into Eq.~(\ref{firstintg}), we arrive at $2\tilde{\lambda} r=c_{0}( 4-c_{0}r)$,
$\tilde{p}r^{2}=-2 c_{0}$ and
$\eta_{0} = -1/r \neq 0$.

\subsubsection{Biconcave Discoid}

\begin{figure}[pth!]
\centerline{\includegraphics[width=9cm]{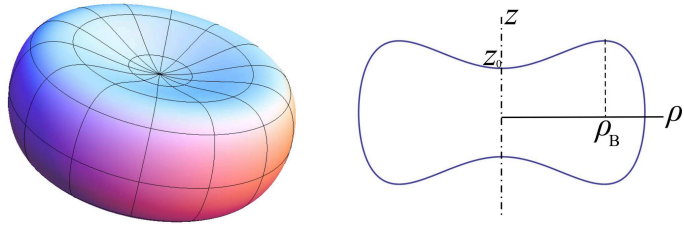}}
\caption{Biconcave discoid (left) and its generation curve (right).\label{figbiconcv}}
\end{figure}

For $0<c_0\rho_B<\re$, the parameter
equation
\begin{equation}\left\{\begin{array}{l}\sin\psi=-c_0\rho\ln(\rho/\rho_B)\\
z=z_0+\int_0^\rho \tan\psi d\rho
\end{array}\right.\label{solutionbicon}\end{equation}
corresponds to a planar curve shown in Fig.~\ref{figbiconcv}. Substituting it into Eq.~(\ref{firstintg}),
we have $\tilde{p}=0$, $\tilde{\lambda}=0$, and
$\eta_0=2c_0 \neq 0$. That is, a biconcave
discoid generated by revolving this planar curve around
$z$-axis can satisfy the shape equation of vesicles. This result can give a good explanation to
the shape of human red blood cells under normal physiological
conditions \cite{EvansMR72,NaitoPRE93,NaitoPRE96}.

A small comment is that $\eta_0=2c_0 \neq 0$ reflects the singularity at two poles of the biconcave discoid. Whether does this singularity exist in real red blood cells? What is the biological meaning of this singularity? Or does there exist a normal solution to the shape equation that can also explain the shape of red blood cells? These open questions need further discussions.

\subsubsection{Unduloid-Like and Cylinder-Like Surfaces}
If we only concern solutions to the shape equation, two cases are also widely discussed. The first case \cite{NaitoPRL95,MladenovEPJB02} is an axisymmetric surface generated by planar curve satisfying
\begin{equation}
\sin\psi = \frac{1}{\rho_m c_0}\left(\frac{\rho}{\rho_m}+\frac{\rho_m}{\rho}\right)- \sqrt{\frac{4}{\rho_m^2 c_0^2}-2},~~(0<\rho_m c_0<4/3).\label{UnduloidLk}
\end{equation}
The generated surface abides by the shape equation with $\tilde{p}=-2 c_0 \rho_m^4$, $\tilde{\lambda}=2/\rho_m^2-c_0^2/2$, and $\eta_0 =2c_0-3/c_0\rho_m^2$. This surface has the unduloid-like shape but with nonconstant mean curvature.

The second one is a cylinder-like surface generated by a planar curve translating along the normal of the plane. If we denote the curvature of the curve as $\kappa$, then the geometric quantities of the generated surface can be expressed as $2H=-\kappa$, $K=0$, and $\nabla^2(2H)=-\ddot{\kappa}$, where the `dot' above $\kappa$ represents the derivative with respect to the arc length of the curve. Thus shape equation (\ref{shape-closed}) degenerates into \cite{ZhangOYPRE96,GuvenPRE20022D,MladenovJPA08}:
\begin{equation}\tilde{p}+\bar{\lambda}
\kappa+\kappa^3/2-\ddot{\kappa}=0\label{shape-cylsurf}\end{equation}
with $\bar{\lambda}=\tilde{\lambda}+c_0^2/2$. The above equation is integrable, which results in
\begin{equation}\dot{\kappa}^2=\xi_0+2\tilde{p}\kappa+\bar{\lambda}\kappa^2-\kappa^4/4\label{shape-clsint}\end{equation}
with an integral constant $\xi_0$. This equation can be further solved in terms of Elliptic functions \cite{ZhangOYPRE96,GuvenPRE20022D,MladenovJPA08,Zhouxh2010}.

It is necessary to note that these two cases do not correspond to real vesicles because they are not closed surfaces.

\section{Lipid Membranes with Free Edges\label{sec-openmem}}

The opening-up process of liposomal membranes by talin \cite{Hotani98} was observed, which gives rise to the study of equilibrium equation and boundary conditions of lipid membranes with free exposed edges. This problem was theoretically investigated by Capovilla \textit{et al.} \cite{CapovillaPRE02} and Tu \textit{et al.} \cite{TuPRE03} in terms of different methods. In this section, we will present these theoretical results and subsequent advancements.

\begin{figure}[pth!]
\centerline{\includegraphics[width=7cm]{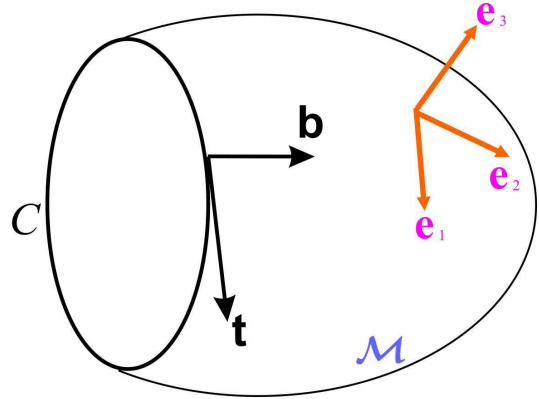}}
\caption{An open smooth surface ($\mathcal{M}$) with a boundary curve ($C$). $\{\mathbf{e}_1,\mathbf{e}_2,\mathbf{e}_3\}$ forms the frame at some point on the surface. $\{\mathbf{t},\mathbf{b},\mathbf{e}_3\}$ also forms the right-handed frame for the point in $C$ such that $\mathbf{t}$ is the tangent of $C$ and $\mathbf{b}$ points to the side that the surface is located in. \label{figopenm}}
\end{figure}

\subsection{Shape Equation and Boundary Conditions to Describe Equilibrium Configurations}

As shown in Fig.~\ref{figopenm}, a lipid membrane with a free edge can be expressed as an open smooth surface ($\mathcal{M}$) with a boundary curve ($C$) in geometry. Because the free exposed edge is energetically unfavorable, we assign the line tension (energy cost per length) to be $\gamma>0$. Then the free energy functional that we need to minimize can be expressed as
\begin{equation}F = \int_\mathcal{M} \left[ \frac{k_c}{2} (2H +c_0)^2 + \bar{k} K \right] \rd A +\lambda A +\gamma L ,\label{eq-frenergyn2}\end{equation}
where $L$ is the total length of the free edge.

By using the variational method introduced in Sec.~\ref{sec-prelim}, we can arrive at a shape equation \cite{TuPRE03}
\begin{equation}(2H+c_{0})(2H^{2}-c_{0}H-2K)-2\tilde\lambda H+\nabla
^{2}(2H) =0, \label{eq-openm}\end{equation}
and three boundary conditions \cite{TuPRE03}
\begin{eqnarray}
&&\left. \lbrack (2H+c_{0})+\tilde{k}\kappa_n]\right\vert _{C} =0,\label{bound1} \\
&&\left. \lbrack -2{\partial H}/{\partial\mathbf{b}}+\tilde\gamma
\kappa_n+\tilde{k}\dot{\tau}_g]\right\vert _{C} =0,\label{bound2}\\
&&\left. \lbrack (1/{2})(2H+c_{0})^{2}+\tilde{k}K+\tilde\lambda
+\tilde\gamma \kappa_{g}]\right\vert _{C}=0,\label{bound3}
\end{eqnarray}
where $\tilde{\lambda}\equiv\lambda/k_c$,
$\tilde{k}\equiv\bar{k}/k_c$, and $\tilde{\gamma}\equiv\gamma/k_c$ are the reduced surface tension, reduced bending modulus, and reduced line tension, respectively. $\kappa_n$, $\kappa_g$, and $\tau_g$ are the normal
curvature, geodesic curvature, and geodesic torsion of the boundary curve, respectively. The `dot' represents the derivative with respect to the arc length of the edge. Equation~(\ref{eq-openm}) expresses
the normal force balance of the membrane. Equations (\ref{bound1})--(\ref{bound3}) represent the force and moment balances at each point in
curve $C$. Thus, in general, the above four
equations are independent of each other and available for an open membrane with several edges.

\begin{figure}[pth!]
\centerline{\includegraphics[width=7cm]{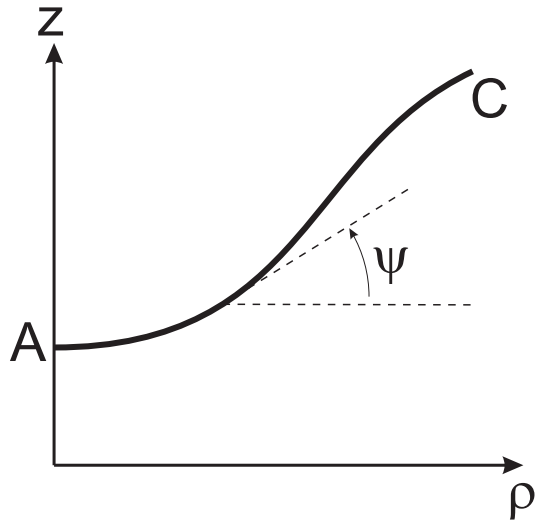}}
\caption{\label{figoutline}
Outline of an axisymmetric open surface. Each open surface can be generated by a
planar curve AC rotating around z axis. $\psi$ is the angle between
the tangent line and the horizontal plane.}
\end{figure}

Now we consider axisymmetric membranes. When a planar curve AC shown in Fig.\ref{figoutline} revolves around
$z$ axis, an axisymmetric surface is generated. Let $\psi$ represent
the angle between the tangent line and the horizontal plane. Each
point in the surface can be expressed as vector
$\mathbf{r}=\{\rho\cos \phi,\rho\sin \phi,z(\rho)\}$, where $\rho$
and $\phi$ are the radius and azimuth angle that the point corresponds
to. Introduce a notation $\sigma$ such that $\sigma =1$ if
$\mathbf{t}$ is parallel to $\partial\mathbf{r} /\partial \phi$,
and $\sigma =-1$ if $\mathbf{t}$ is antiparallel to
$\partial\mathbf{r} /\partial \phi$ in the boundary curve generated
by point C. The above equations (\ref{eq-openm})--(\ref{bound3}) are
transformed into
\begin{eqnarray}
(h-c_{0})\left(\frac{h^{2}}{2}+\frac{c_{0}h}{2}-2K\right)-\tilde{\lambda}
h+\frac{\cos \psi }{\rho}(\rho\cos \psi h')'=0,\label{sequilib}
\\
\left[h-c_{0}+\tilde{k}{\sin \psi }/{\rho}\right]_C=0,\label{sbound1}\\
\left[-\sigma\cos \psi h'+\tilde{\gamma}{\sin \psi
}/{\rho}\right]_C=0,\label{sbound2}\\
\left[\frac{1}{2}(h-c_0)^2+\tilde{k}K+\tilde{\lambda}-\sigma\tilde{\gamma}
\frac{\cos \psi }{\rho}\right]_C=0,\label{sbound3}
\end{eqnarray}
with $h\equiv {\sin \psi }/{\rho}+(\sin\psi)'$ and $K\equiv{\sin \psi
}(\sin\psi)'/{\rho}$. The `prime' represents the derivative with
respect to $\rho$.

Shape equation (\ref{sequilib}) is integrable, which reduces to a second order differential equation
\begin{equation}\cos\psi h'
+(h-c_{0}) \sin\psi\psi^{\prime}-\tilde{\lambda} \tan\psi+\frac{\eta_{0}}{\rho\cos\psi}-\frac{\tan\psi}
{2}(h-c_{0})^{2} =0\label{firstintg2}\end{equation} with an integral
constant $\eta_{0}$ \cite{TuJCP2010}. This equation is equivalent to Eq.~(\ref{firstintg}) with zero osmotic pressure. The configuration of an axisymmetric open lipid membrane should satisfy
shape equation (\ref{firstintg2}) and
boundary conditions (\ref{sbound1})--(\ref{sbound3}). In particular,
the points in the boundary curve should satisfy not only the
boundary conditions, but also shape equation (\ref{firstintg2})
because they also locate in the surface. That is,
Eqs.~(\ref{sbound1})-(\ref{sbound3}) and (\ref{firstintg2}) should be
compatible with each other in the edge. Substituting
Eqs.~(\ref{sbound1})-(\ref{sbound3}) into (\ref{firstintg2}), we
derive the compatibility condition \cite{TuJCP2010} to be
\begin{equation}\eta_{0}=0.\label{compat-cond}\end{equation} It is a necessary (not sufficient)
condition for existence of axisymmetric open membranes. Under this condition, the shape equation
is reduced to
\begin{equation}\cos\psi h'
+(h-c_{0}) \sin\psi\psi^{\prime}
  -\tilde{\lambda} \tan\psi-\frac{\tan\psi}%
{2}(h-c_{0})^{2} =0,\label{newshapeq}\end{equation}
while three boundary conditions are reduced to two equations, i.e. Eqs.~(\ref{sbound1}) and (\ref{sbound3}).

\subsection{Theorems of Non-Existence}

Now our task is to find analytic solutions that satisfy both the shape equation and boundary conditions.
An obvious but trivial one is a circular disk with radius $R$. In this case, Eqs.~(\ref{eq-openm})--(\ref{bound3}) degenerate to
\begin{equation}\tilde{\lambda} R+ \tilde{\gamma} =0.\end{equation}

Can we find nontrivial analytic solutions? We will prove several theorems of non-existence in this subsection, which imply that it is almost hopeless to find nontrivial analytic solutions.

\begin{theorem}    \label{thm-sphere}
There is no open membrane being a part of a spherical vesicle.
\end{theorem}

\textbf{Proof}. For a sphere with radius $R$, we can calculate $H=-1/R$, $\kappa_n = -1/R$ and $\tau_g=0$ in terms of Eq.~(\ref{geodisicc}) because $a=c=-1/R$ and $b=0$ for a sphere. Boundary condition (\ref{bound2}) cannot be abided by. Thus an open membrane cannot be a part of a spherical vesicle.

\begin{theorem} \label{thm-cyld}
There is no open membrane being a part of a cylindrical surface.
\end{theorem}

\textbf{Proof}. For any line element on the surface of a cylinder with radius $R$, we can calculate $\kappa_n = -\cos^2\theta/R$ from Eq.~(\ref{geodisiccprfm}) where $\theta$ is the angle between the line element and the circumferential direction. Additionally, $H=-1/R$ is a constant. If $\tilde{k}=0$, then boundary condition (\ref{bound2}) results in $\kappa_n =0$, that is $\theta =\pi/2$. The line along this direction is not a closed curve, and so cannot be as an edge of a membrane. If $\tilde{k}\neq 0$, then boundary condition (\ref{bound1}) results in $\kappa_n=(c_0-1/R)/\tilde{k}$, which implies $\theta$ should be a constant.
The unique closed curve is a circle, i.e. $\theta=0$ and $\kappa_n=-1/R$. But $\tau_g=0$ if $\theta=0$, then contradicts with boundary condition (\ref{bound2}). Thus an open membrane cannot be a part of a spherical surface.

\begin{theorem}    \label{thm-constH}
There is no open membrane being a part of a curved surface with constant mean curvature.
\end{theorem}

\textbf{Proof}. Two special surfaces (sphere and cylinder) with constant mean curvature are discussed in the above theorems. Now we only need to investigate surfaces with constant $H$ but nonconstant $K$. From shape equation of open membranes, we derive two possible cases: (i) $H=-c_0/2\neq 0$ and $\tilde{\lambda} =0$; (ii) $H=c_0=0$ and $\tilde{\lambda} \neq 0$.

In the former case, if $\tilde{k}=0$, then boundary conditions (\ref{bound2}) and (\ref{bound3}) result in $\kappa_n =\kappa_g =0$. Thus the curvature of boundary curve is $\kappa =\sqrt{\kappa_n^2+ \kappa_g^2}=0$. That is, this curve is a straight line which is not closed curve. If $\tilde{k}\neq 0$, then boundary conditions (\ref{bound1}) and (\ref{bound2}) results in $\kappa_n=0$ and $\tau_g = \mathrm{constant}$. Using Eq.~(\ref{geodisiccprfm}), we derive the principal curvatures for the points in the curve are constant. Then Eq.~(\ref{bound3}) requires $\kappa_g= \mathrm{constant}$. That is, the curvature and torsion are constant in the curve. The unique closed curve is a circle. But $\tau_g =0$ for a circle. Let $c_1$ and $c_2$ represent the two principal curvatures, $\theta$ is the angle between the tangent of the curve and one principal direction at each point in the curve. Then we have two equations: $\kappa_n=c_1 \cos^2\theta +c_2 \sin^2\theta =0$ and $\tau_g=(c_2-c_1)\sin\theta\cos\theta=0$. Substituting these two equations and $K=c_1c_2$ into Eq.~(\ref{bound3}), we obtain $\kappa_g=0$. Then $\kappa =\sqrt{\kappa_n^2+ \kappa_g^2}=0$, which contradicts with the preassumption of a circle.

In the latter case, $H=c_0=0$, similar to the proof in the former one, it also leads to a contradiction. Thus there is no open membrane being a part of a curved surface with constant mean curvature.

\begin{theorem}    \label{thm-wlms}
There is no open membrane being a part of Willmore surface.
\end{theorem}

\textbf{Proof}. Let us consider the scaling transformation $\mathbf{r}\rightarrow \Lambda\mathbf{r}$, where the vector $\mathbf{r}$ represents the position of each point in the membrane and $\Lambda$ is a scaling parameter \cite{CapovillaJPA02}. Under this transformation, we have $A\rightarrow \Lambda^2 A$, $L\rightarrow \Lambda L$, $H\rightarrow \Lambda^{-1} H$, and $K\rightarrow \Lambda^{-2} K$. Thus, Eq.~(\ref{eq-frenergyn2}) is transformed into
\begin{eqnarray}F(\Lambda)&=&\int_{\mathcal{M}} [(k_c/2)(2H)^2 +\bar{k}K] \rd A\nonumber\\ &+& 2k_c c_0 \Lambda \int_{\mathcal{M}} H \rd A+(\lambda+k_c c_0^2 /2)\Lambda^2 A +\gamma \Lambda L. \label{eq-frenergyn3}\end{eqnarray}

The equilibrium configuration should satisfy $\partial F/\partial\Lambda =0$ when $\Lambda=1$. Thus we obtain
\begin{equation}2 c_0 \int_{\mathcal{M}} H \rd A+(2\tilde\lambda+c_0^2) A +\tilde\gamma L=0.\label{constraitg}\end{equation}
This equation is an additional constraint for open membranes.

Willmore surfaces satisfy the special form of Eq.~(\ref{shape-closed}) with vanishing $\tilde\lambda$ and $c_0$ \cite{Willmore82}. Because $\tilde\gamma L>0$, thus the constraint (\ref{constraitg}) cannot be satisfied when $\tilde\lambda=0$ and $c_0=0$. That is, there is no open membrane being a
part of Willmore surface.

\begin{figure}[pth!]
\centerline{\includegraphics[width=12cm]{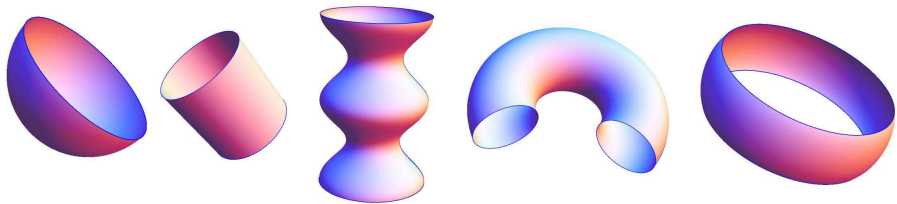}}
\caption{\label{fig-impopm}
Schematics of several impossible open membranes with free edges: parts of sphere, cylinder, unduloid, torus, biconcave discodal surface (from left to right).}
\end{figure}

\textbf{Corollary} \textit{There is no open membrane being a
part of $\sqrt{2}$ torus.}

\textbf{Proof}. Substituting $\sqrt{2}$ torus into Eq.~(\ref{eq-openm}), we obtain $c_0=0$ and $\tilde{\lambda}=0$. That is, $\sqrt{2}$ torus is a Willmore surface. In terms of theorem \ref{thm-wlms}, we arrive at this corollary.

\begin{theorem}  \label{thm-bcdisk}
There is no axisymmetric open membrane being a
part of a biconcave discodal surface generated by a planar curve
expressed by $\sin\psi=-c_0 \rho \ln(\rho/\rho_B)$.
\end{theorem}

\textbf{Proof}. A biconcave discodal surface \cite{NaitoPRE93,NaitoPRE96}
generated by a planar curve expressed by $\sin\psi=-c_0 \rho \ln(\rho/\rho_B)$ with non-vanishing constants $c_0$ and
$\rho_B$. To avoid the singularity at two poles,
we may dig two holes around the poles.
Substituting this equation into shape equation (\ref{firstintg2}),
we obtain $\tilde{\lambda}=0$ and $\eta_0 = 2 c_0$.
That is, the biconcave discodal surface can be a solution to the
shape equation. However, $\eta_0 = 2c_0 \neq 0$ contradicts to compatibility
condition (\ref{compat-cond}). Thus there is no axisymmetric open membrane being a
part of this biconcave discodal surface.

In Fig.~\ref{fig-impopm}, we show several impossible open membranes with free edges in terms of the above theorems. These theorems suggest that it is hopeless to find exactly analytic solutions to the shape equation and boundary conditions of open lipid membranes. Quasi-exact solutions or numerical simulations are highly appreciated.

\subsection{Quasi-Exact Solutions \label{sbusec-qesolu}}
\begin{figure}[pth!]
\centerline{\includegraphics[width=11cm]{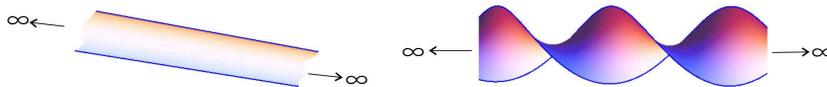}}
\caption{\label{figsemexact}
Schematics of two quasi-exact solutions: straight stripe along axial direction of cylinder (left) and twist ribbon which is a part of a minimal surface (right).}
\end{figure}

Here the quasi-exact solution is defined as a surface with free edge(s) such that the points on that surface exactly satisfy the shape equation, and most of points in the edge(s) abide by boundary conditions. In fact, the proves to theorems \ref{thm-cyld} and \ref{thm-constH} implies two possible solutions as shown in Fig.~\ref{figsemexact}. One is a straight stripe along the axial direction of cylinder, another is a twist ribbon which is a part of a minimal surface ($H$=0).

Let we consider a long enough straight stripe along the axial direction of cylinder that satisfies shape equation (\ref{eq-openm}), that is, $\tilde{\lambda}=(1-c_0^2R^2)/2R^2$. The long enough configuration ensure us to omit the boundary of two ends. The lateral edges are straight lines which have $\kappa_n=\kappa_g=\tau_g =0$. Thus boundary conditions (\ref{bound1}) and (\ref{bound2}) are trivial. The third boundary condition results in $\tilde{\lambda}=(1-c_0R)^2/2R^2$. Thus we arrive at $\tilde{\lambda}=0$ and $R=1/c_0$. That is, a long enough straight stripe along the axial direction of cylinder with $R=1/c_0$ is a quasi-exact solution.

Next, a twisted ribbon with pitch $T$ and width $2u_0$ can be expressed as vector form
$\{u\cos\varphi,u\sin\varphi,\alpha\varphi\}$ with $|u|\leq u_0$,
$|\varphi|<\infty$ and $|\alpha|=T/2\pi$.
From simple calculations, we have
\begin{equation}H=0,~ K=-\alpha^{2}/(u^{2}+\alpha ^{2})^2\label{twistsolut}\end{equation} for the points on the surface, and
\begin{align}
\kappa_n&=0, K=-\alpha^{2}/(u_0^{2}+\alpha^{2})^{2}\notag\\
\kappa_g&=u_0/(u_0^{2}+\alpha^{2})\label{kpataubdtwist}\\
\tau_g&=\alpha/(u_0^{2}+\alpha^{2})\notag
\end{align}
for the points on the edges.

It is easy to see that equation (\ref{twistsolut}) can satisfy shape equation (\ref{eq-openm}) when $c_0=0$. Then equation (\ref{kpataubdtwist}) naturally validates boundary conditions (\ref{bound1}) and (\ref{bound2}). The last boundary condition (\ref{bound3}) leads to $\tilde{\lambda}=[\tilde{k}\alpha^2 -\tilde{\gamma}u_0(u_0^2+\alpha^2)]/(u_0^2+\alpha^2)^2$, which can be satisfied by proper parameters $\tilde{\lambda}$, $\tilde{k}$, $\tilde{\gamma}$, $u_0$ and $\alpha$. That is, the twist ribbon is indeed a quasi-exact solution.

\section{Chiral Lipid Membranes\label{sec-chiralm}}

In fact, our above discussions only concern lipid membranes where lipid molecules are in Smectic A phase.
In this phase, lipid molecules almost point to the normal direction of the membrane surface. However, there are also many kinds of chiral lipids in cell membranes. At body temperature, chiral lipids usually form Smectic $\mathrm{C}^\ast$ phase. They are tilting from the normal direction in a constant angle. It is necessary to develop Helfrich's spontaneous curvature model introduced in Sec.~\ref{subs-helfmd} to cover the Smectic $\mathrm{C}^\ast$ phase. Based on symmetric argument or Frank energy in the theory of liquid crystal, many theoretical models and results were achieved \cite{Helfrich88,oy-liuprl90,oy-liupra90,Nelson92,Selinger96,SelingerJPCB96,Komuraoyprl98}.
These theoretical models contain much complicated terms and many parameters, which make it is impossible to derive the exact governing equations for describing equilibrium configurations of chiral lipid membranes. Here we will discuss a simplified version proposed by the present author and Seifert \cite{Tupre2007}. It is found that this concise theory can still explain most of experimental phenomena.

\subsection{Constructing the Free Energy}

The free energy density for a chiral lipid membrane are supposed to consist of the following contributions.

(i) The bending energy per area is still taken as Helfrich's
form (\ref{helfrichedens}). That is, we neglect the anisotropic effect of lipid molecules' tilting on the bending moduli.

\begin{figure}[pth!]
\centerline{\includegraphics[width=8cm]{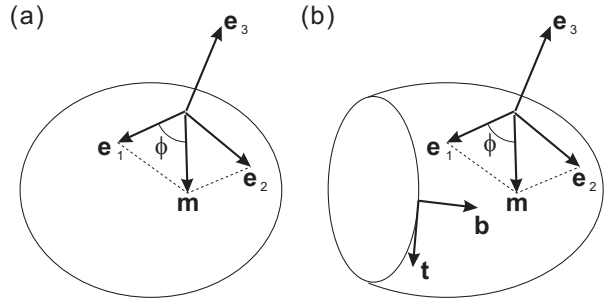}}
\caption{\label{figframeclm}
Right-handed orthonormal frame
$\{\mathbf{e}_1,\mathbf{e}_2,\mathbf{e}_3\}$ at any point in a
surface where $\mathbf{e}_3$ is the normal vector of the surface.
(a) Surface without boundary curve. (b) Surface with boundary curve
where $\mathbf{t}$ is the tangent vector of the boundary curve, and
$\mathbf{b}$, in the tangent plane of the surface, is perpendicular
to $\mathbf{t}$. (Reprint from Ref.~\cite{Tupre2007})}
\end{figure}

(ii) The energy per area originating from the chirality of tilting
molecules has the form \cite{oy-liuprl90}
\begin{equation}f_{ch}=-h\tau_{\mathbf{m}},\label{chiralen}\end{equation}
where $h$ reflects the strength of molecular chirality. Without losing the generality, here we only
discuss the case of $h>0$. $\tau_{\mathbf{m}}$ is the
geodesic torsion along the unit vector $\mathbf{m}$ at each point.
Here $\mathbf{m}$ represents the projected direction of the lipid
molecules on the membrane surface. If we take a right-handed
orthonormal frame $\{\mathbf{e}_1,\mathbf{e}_2,\mathbf{e}_3\}$ as shown in
Fig.~\ref{figframeclm}, $\mathbf{m}$ can be expressed as
$\mathbf{m}=\cos\phi\mathbf{e}_1+\sin\phi\mathbf{e}_2$, where $\phi$
is the angle between $\mathbf{m}$ and $\mathbf{e}_1$. then geodesic torsion $\tau _{\mathbf{m}}$ and normal curvature $\kappa _{\mathbf{m}}$ along $\mathbf{m}$ can be expressed as
the similar form of Eq.~(\ref{geodisicc}).

(iii) The energy per area due to the orientational variation is taken as
\begin{equation}f_{ov}=(k_{f}/2)[(\nabla\times\mathbf{m})^{2}+(\nabla\cdot\mathbf{m})^{2}],\end{equation}
where $k_f$ is a constant in the dimension of energy. This is the
simplest term of energy cost due to tilting order invariant under
the coordinate rotation around the normal of the membrane surface.
By defining a spin connection field $\mathbf{S}$ such that
$\nabla\times\mathbf{S}=K$, one can derive $(\nabla\times\mathbf{m}) ^{2}+(
\nabla\cdot\mathbf{m})^{2}=(\nabla\phi-\mathbf{S})^{2}$
through simple calculations \cite{Nelson87}.

The total free energy density adopted in the present paper,
$G=f_{c}+f_{ch}+f_{ov}$, has the following concise form:
\begin{equation}
G=\frac{k_{c}}{2}(2H+c_{0})^{2}+\bar{k}K-h\tau_{\mathbf{m}}+\frac{k_{f}}{2}\mathbf{v}
^{2},\label{energy2}
\end{equation}
with $\mathbf{v}\equiv\nabla\phi-\mathbf{S}$. This special form
might arguably be the most natural and concise construction
including the bending, chirality and tilting order, for the given
vector field $\mathbf{m}$ and normal vector field $\mathbf{e}_3$.

\subsection{Governing Equations to Describe Equilibrium Configurations}

The free energy for a closed chiral lipid vesicle may be expressed as
\begin{equation}F=\int_{\mathcal{M}} G\rd A +\lambda A+p V,\label{closedFE}\end{equation}
where $A$ is the area of the membrane and $V$ the volume enclosed by the vesicle. $\lambda$ and $p$ are two multipliers to implement area and volume constraints.

Using the variational method mentioned in Sec.\ref{sec-prelim}, we can obtain two
governing equations to describe equilibrium configurations \cite{Tupre2007} as
\begin{equation}
2\tilde{h}(\kappa_{\mathbf{m}}-H)-\tilde{k}_{f}\nabla^{2}\phi=0 \label{EL1}
\end{equation}
and
\begin{eqnarray}
&&\hspace{-0.5cm}2\nabla^{2}H+( 2H+c_{0}) (
2H^{2}-c_{0}H-2K)-2\tilde{\lambda} H+\tilde{p}\nonumber\\
&&\hspace{-0.5cm}+\tilde{h}[  \nabla\cdot( \mathbf{m}\nabla\times\mathbf{m})
+\nabla\times(
\mathbf{m}\nabla\cdot\mathbf{m})]\nonumber\\
&&\hspace{-0.5cm}+\tilde{k}_{f}[( \kappa_{\mathbf{v}}-H) \mathbf{v}^{2}-
\nabla\mathbf{v}\colon\nabla\mathbf{e}_3] =0\label{EL2}
\end{eqnarray}
with reduced parameters $\tilde{h}=h/k_c$, $\tilde{k}_{f}={k}_{f}/k_c$, $\tilde{p}=p/k_c$, and $\tilde{\lambda}=\lambda/k_c$. $\kappa_{\mathbf{m}}$ and $\kappa_{\mathbf{v}}$ are the normal
curvature along the directions of $\mathbf{m}$ and $\mathbf{v}$,
respectively. When writing Eq.\,(\ref{EL1}), we have
selected the proper gauge such that $\nabla\cdot\mathbf{S}=0$, or
else $\nabla^{2}\phi$ should be replaced with
$\nabla^{2}\phi-\nabla\cdot\mathbf{S}$. Additionally, we do not consider singular points for closed vesicles
different from toroidal topology.

Consider a chiral lipid membrane with a free edge as shown in Fig.~\ref{figframeclm}b.
Its free energy can be expressed as
\begin{equation}F=\int_{\mathcal{M}} G \rd A+\lambda A +\gamma L ,\label{openFE}\end{equation}
where $A$ is the area of the membrane and $L$ the total length of the edge. $\gamma$ represents the line tension of
the edge.

Using the variational method mentioned in Sec.\ref{sec-prelim}, we can obtain the
governing equations to describe equilibrium configurations  of membrane surfaces as
\begin{equation}
2\tilde{h}(\kappa_{\mathbf{m}}-H)-\tilde{k}_{f}\nabla^{2}\phi=0 \label{EL3}
\end{equation}
and
\begin{eqnarray}
&&2\nabla^{2}H+( 2H+c_{0}) (
2H^{2}-c_{0}H-2K)-2\tilde{\lambda} H\nonumber\\
&&+\tilde{h}[  \nabla\cdot( \mathbf{m}\nabla\times\mathbf{m}) +\nabla\times(
\mathbf{m}\nabla\cdot\mathbf{m})]\nonumber\\
&&+\tilde{k}_{f}[( \kappa_{\mathbf{v}}-H) \mathbf{v}^{2}-
\nabla\mathbf{v}\colon\nabla\mathbf{e}_3] =0.\label{EL4}
\end{eqnarray}
Simultaneously, the boundary conditions obeyed by the free edge are
derived as \cite{Tupre2007}:
\begin{eqnarray}&&\hspace{-0.88cm}v_{b}=0,\label{BC1}\\
&&\hspace{-0.88cm}(1/2)(2H+c_{0})^{2}+\tilde{k}K-\tilde{h}\tau_{\mathbf{m}}+(\tilde{k}_{f}/2)\mathbf{v}^2+\tilde{\lambda}+\tilde{\gamma}\kappa_{g}=0,\label{BC2}\\
&&\hspace{-0.88cm}(2H+c_{0})+\tilde{k}\kappa_{n}-(\tilde{h}/2)\sin2\bar{\phi}=0,\label{BC3}\\
&&\hspace{-0.88cm}\tilde{\gamma}\kappa_{n}+\tilde{k}\dot{\tau}_{g}-2\partial H/\partial\mathbf{b}-\tilde{h}(v_{t}+\dot{\bar{\phi}})\sin2\bar{\phi}
+\tilde{k}_{f}\kappa_{n}v_{t}=0,\label{BC4}\end{eqnarray} where
$\kappa_{n}$, $\tau_{g}$ and $\kappa_{g}$ are the normal curvature,
geodesic torsion, and geodesic curvature of the boundary curve
(i.e., the edge), respectively. $v_b$ and $v_t$ are the components of
$\mathbf{v}$ in the directions of $\mathbf{b}$ and $\mathbf{t}$,
respectively. The `dot' represents the derivative with
respect to arc length parameter $s$. $\bar\phi$ is the angle between $\mathbf{m}$ and
$\mathbf{t}$ at the boundary curve. Equations (\ref{BC1})--(\ref{BC4}) describe the force and moment balance
relations in the edge. Thus they are also available for a chiral lipid membrane with
several edges.

\subsection{Solutions and Corresponding Configurations}
Now we will present some analytic solutions to the governing equations of chiral lipid membranes.

\subsubsection{Sphere}
For spherical vesicles of chiral lipid molecules with radius $R$,
$\tau_{\mathbf{m}}$ is always vanishing because $a=c=1/R$ and $b=0$. Thus the
free energy (\ref{closedFE}) is independent of the molecular
chirality and permits the same existence probability of left- and right-handed
spherical vesicles. This is uninteresting case in practice.

\subsubsection{Cylinder}

Here we consider a long enough cylinder with radius $R$ such that its two ends can be neglected. The cylinder can be parameterized by two variables $s$ and $z$ which are the arc length along the circumferential direction and coordinate along axial direction, respectively. Let $\phi$ be the angle between $\mathbf{m}$ and the circumferential direction. Then Eqs.\,(\ref{EL1}) and (\ref{EL2}) are transformed
into \cite{Tupre2007}:
\begin{equation}\tilde{k}_{f}(\phi_{ss}+\phi_{zz})+(\tilde{h}/R)\cos2\phi=0,\label{stripe01}\end{equation}
and
\begin{eqnarray}&&\hspace{-0.4cm}\tilde{h}[ 2(
\phi_{z}^{2}-\phi_{s}^{2}+\phi_{sz})\sin2\phi+(
\phi_{ss}-\phi_{zz}+4\phi_{z}\phi_{s})\cos2\phi]+\tilde{\lambda}/R\nonumber\\&&\hspace{-0.4cm}+\tilde{p}+(
c_{0}^{2}-1/R^{2}) /2R+\tilde{k}_{f}[ (
\phi_{z}^{2}-\phi_{s}^{2})/2R+\phi_{sz}/R]=0.\label{stripe02}\end{eqnarray}
where the subscripts $s$ and $z$ represent the partial derivatives
respect to $s$ and $z$, respectively.

\begin{figure}[pth!]
\centerline{\includegraphics[width=10cm]{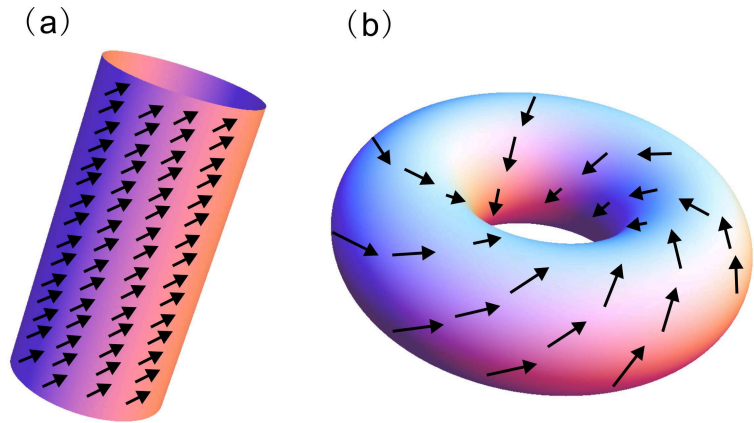}}
\caption{\label{fig-solutchm}
Two possible chiral lipid membranes: (a) Cylinder with uniform tilting state; (b) Torus with uniform tilting state.}
\end{figure}

It is not hard to see that $\phi =\pi/4$ and $2\tilde{p}R^3+2\tilde{\lambda}
R^2-1+c_0^2R^2=0$ can satisfy the above two equations. Thus a cylinder shown in Fig.~\ref{fig-solutchm}a with uniform tilting state (tilting angle $\phi =\pi/4$) is a solution.

\subsubsection{Torus}

A torus is a revolution surface generated by a cycle with radius
$r$ rotating around an axis in the same plane of the cycle as
shown in Fig.~\ref{figtorus}. It can be expressed as
vector form
$\{(R+r\cos\varphi)\cos\theta,(R+r\cos\varphi)\sin\theta,r\sin\varphi\}$.
Equation\,(\ref{EL1}) is
transformed into \cite{Tupre2007}:
\begin{equation}\frac{1}{\nu+\cos\varphi}\frac{\partial^2\phi}{\partial \theta^2}+\frac{\partial}{\partial\varphi}\left[(\nu+\cos\varphi)  \frac{\partial\phi}{\partial\varphi}\right]-\frac{\nu \tilde{h}r}{\tilde{k}_{f}}\cos2\phi=0,\label{phitorus}\end{equation}
where $\phi$ is the angle between $\mathbf{m}$ and the latitude of
the torus, while $\nu\equiv R/r$ is the ratio between two
generated radii of the torus.

The uniform tilting state ($\phi=-\pi/4$) satisfies
Eq.~(\ref{phitorus}) and makes $-\int
h\tau_\mathbf{m}\rd A$ to take the minimum. With
$\phi=-\pi/4$, Eq.\,(\ref{EL2}) is transformed into \cite{Tupre2007}:
\begin{eqnarray}
&&~~~\hspace{-0.5cm}(2-\tilde{k}_f)/{\nu^2}+(c_0^2r^2-1)+2(\tilde{p}r+\tilde{\lambda})r^2\nonumber\\
&&\hspace{-0.5cm}+[(4 c_0^2r^2-4 c_0r-2 \tilde{h} r+8\tilde{\lambda} r^2+6\tilde{p}r^3)/{\nu}]\cos\varphi\nonumber\\
&&\hspace{-0.5cm}+[(5 c_0^2r^2-8 c_0r-4 \tilde{h} r+10\tilde{\lambda} r^2+3 \tilde{k}_f+6\tilde{p}r^3)/\nu^2]\cos^2\varphi\nonumber\\
&&\hspace{-0.5cm}+[(2 c_0^2r^2-4 c_0r-2 \tilde{h}
r+4\tilde{\lambda} r^2+2
\tilde{k}_f+2\tilde{p}r^3)/\nu^3]\cos^3\varphi=0.\label{toruseqchm}
\end{eqnarray}
Because $\nu$ is finite for a torus, then the above equation
holds if and only if the coefficients of
$\{1,\cos\varphi,\cos^2\varphi,\cos^3\varphi\}$ vanish. It follows
that $2\tilde{\lambda} r^{2}=( 4r c_{0}-r^{2}c_{0}^{2})
-3\tilde{k}_{f}+2\tilde{h}r$, $\tilde{p}r^{3}=2\tilde{k}_{f}-2r c_{0}-\tilde{h}r$ and
\begin{equation}\nu=\sqrt{(2-\tilde{k}_f)/(1-\tilde{k}_f)}.\label{torusrad}\end{equation}

Thus a torus with uniform tilting state as shown in Fig.~\ref{fig-solutchm}b is an exact solution to governing equations of chiral lipid vesicles. The ratio of two generation radii satisfies Eq.~(\ref{torusrad}), which increases with $\tilde{k}_f$. Especially, $\nu=\sqrt{2}$ for
$\tilde{k}_f=0$, which leads to the $\sqrt{2}$ torus of non-tilting lipid
molecules \cite{oypra90}. Since this kind of torus was
observed in the experiment \cite{MutzPRA91}, tori
with $\nu>\sqrt{2}$ for $0<\tilde{k}_f<1$ might also be observed in some
experiments on chiral lipid membranes.

\subsubsection{Twisted Ribbons}

\begin{figure}[pth!]
\centerline{\includegraphics[height=7cm]{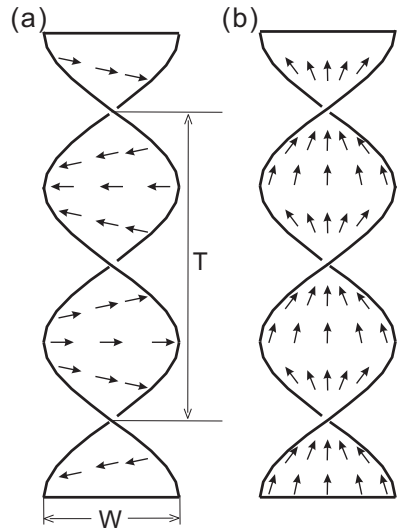}}
\caption{\label{twistrbfig}
Long enough twisted ribbons with lipid molecules in different tilting states: (a) $\mathbf{m}$ is perpendicular to the edges; (b)
$\mathbf{m}$ parallels the edges. Arrows represent the projected
directions $\{\mathbf{m}\}$ of the tilting molecules on the ribbons'
surface. (Reprint from Ref.~\cite{Tupre2007})}
\end{figure}

Here we consider a quasi-exact solution for the governing equations to describing equilibrium configurations of chiral lipid membranes with free edges. Two long enough twisted ribbons with lipid molecules in different tilting states are shown in
Fig.~\ref{twistrbfig}. Similar to Sec.\ref{sbusec-qesolu}, a twist ribbon can be expressed as vector form
$\{u\cos\varphi,u\sin\varphi,\alpha\varphi\}$ with $|u|\leq W/2$,
$|\varphi|<\infty$ and $|\alpha|=T/2\pi$. Equation\,(\ref{EL3}) is transformed into \cite{Tupre2007}:
\begin{equation}\tilde{k}_{f}\left(\phi_{uu
}+\frac{u\phi_{u}+\phi_{\varphi\varphi}}{u^{2}+\alpha^{2}}\right)
+\frac{2\tilde{h}\alpha\sin2\phi}{u^{2}+\alpha^{2}}=0.\label{eqtwistr1}
\end{equation} where $\phi$ is the angle between $\mathbf{m}$
and the horizontal.

If we only consider the uniform tilting state, the above equation
requires $\phi=0$ or $\pi/2$. It is easy to see that $\phi=0$
minimizes $-h\int\tau_\mathbf{m}\rd A$ for $\alpha<0$ while
$\phi=\pi/2$ minimizes $-h\int\tau_\mathbf{m}\rd A$ for $\alpha>0$ because $\tau_\mathbf{m}=-\alpha \cos2\phi/(u^2+\alpha^2)$ \cite{Tupre2007}. Thus we should take $\phi=0$ for $\alpha<0$
and $\phi=\pi/2$ for $\alpha>0$. The former case corresponds to
Fig.~\ref{twistrbfig}a where $\mathbf{m}$ is perpendicular to the
edges; the latter corresponds to Fig.~\ref{twistrbfig}b where
$\mathbf{m}$ is parallel to the edges. Both for $\phi=0$ and $\pi/2$,
Eq.\,(\ref{EL4}) leads to $c_0=0$ for non-vanishing $\alpha$. Among the
boundary conditions (\ref{BC1})--(\ref{BC4}), only Eq.\,(\ref{BC2})
is nontrivial, which reduces to
\begin{equation}\tilde{\lambda}(1+x^{2})  \alpha^{2}-(\tilde{h}-\tilde{\gamma} x)|\alpha|+\frac{\tilde{k}_{f}x^{2}-2\tilde
{k}}{2(1+x^{2})}=0\label{BC22}\end{equation} with
$x\equiv W/2|\alpha|$. Solutions to this equation exists for proper parameters. Thus, there are indeed twist ribbons in two states as shown in Fig.~\ref{twistrbfig}, they have different chirality and tilting angles.

\section{Summary and Conclusion\label{sec-summary}}

In the above discussions, we have presented some theoretical results on the Geometry of membranes, which include the
surface theory and variational method based on moving frame, the governing equations to describe equilibrium configurations of various lipid structures derived from the variation of free energy functionals, some analytic solutions to these equations and their corresponding configurations. We only focus on the pure theoretical researches and miss all experimental and numerical results related to our topic on which gentle readers may consult
Refs.~\cite{EvansMR72,HotaniJMB84,Baumgart03,LimPNAS02,Wangdu08,DuLWJCP06,YanJpre98,Umedapre05}.

Although many theoretical advancements have been achieved, there are still a lot of challenges waiting for further investigations. Several key open questions among them are listed as follows.

(i) Lipid vesicles of multi-components. Cell membranes contains many kinds of lipids. At body temperature, different kinds of lipids usually separate into several lipid domains. Lipid vesicles with two or several domains have been investigated from experimental and theoretical levels \cite{Baumgart03,YinJBP08,GozdzPRE99,JuelicherPRE96,Wangdu08,TuJPA04,Tu2008jctn}. However, there is still lack of strictly exact solutions to the governing equations \cite{TuJPA04,Tu2008jctn} describing the vesicles with multi-domains.

(ii) Other solutions on the shape equations of lipid membranes. We have only found a few analytic solutions to the governing equations of lipid structures. Whether are there other solutions, in particular to the simplest equations (\ref{shape-closed}) and (\ref{firstintg})? Or can we prove that there is no other analytic form except the solutions that we have mentioned?

(iii) Generalized boundary conditions for open lipid membranes. Although we have investigated the boundary conditions of lipid membranes with free edges, there are still other kind cases, such as confined edges, contact lines, and so on. Can we develop a generalized variational principle covering such cases?

(iv) Non-orientable membranes. All membranes that we have considered are orientable membranes. How can we deal with the non-orientable membranes, such as M\"{o}bius band \cite{Zhouxhup}?

\section*{Acknowledgements}
The author is grateful to the financial supports from the Nature Science Foundation of China (grant no.
10704009), the Foundation of National Excellent Doctoral
Dissertation of China (grant no. 2007B17) and the Fundamental Research Funds for the Central Universities.


\aut{Zhanchun Tu\\
Department of Physics\\
Beijing Normal University\\
Beijing 100875, China\\
{\it E-mail address}:
 {\tt tuzc@bnu.edu.cn}} {}
\label{last}
\end{document}